\documentclass[12pt]{article}

\usepackage{amsfonts,amsmath,amssymb}

\usepackage{tikz}

\newcommand{\CC}{{\mathbb C}}
\newcommand{\ZZ}{{\mathbb Z}}
\newcommand{\RR}{{\mathbb R}}

\newcommand{\PP}{{\mathbb P}}

\newtheorem{thm}{Theorem}[section]

\title{Framed Wilson Operators, Fermionic Strings, and Gravitational Anomaly in 4d}

\author{Ryan Thorngren
\\ {\it University of California, Berkeley, CA}}

\begin{document}

\maketitle

\begin{abstract}

We study gapped systems with anomalous time-reversal symmetry and global gravitational anomaly in three and four spacetime dimensions. These systems describe topological order on the boundary of bosonic Symmetry Protected Topological (SPT) Phases. Our description of these phases is via the recent cobordism proposal for their classification. In particular, the behavior of these systems is determined by the geometry of Stiefel-Whitney classes. We discuss electric and magnetic operators defined by these classes, and new types of Wilson lines and surfaces that sit on their boundary. The lines describe fermionic particles, while the surfaces describe a sort of fermionic string. We show that QED with a fermionic monopole exhibits the 4d global gravitational anomaly and has a fermionic $\pi$-flux.

\end{abstract}

\section{Introduction}

We study gapped systems with internal symmetry $G_0$ by gauging the symmetry and considering the effective theory of the gauge field. This theory is topological in simple situations, making this program tractable. However, if we wish to study some larger symmetry group $G$ which contains space-time symmetries, it is not so clear how to proceed. For instance, one needs to know what it means to gauge $G$, and in particular what sort of object the $G$ gauge field is. This is the question that occupies us here.

One of the primary applications of this program is the classification of Symmetry Protected Topological (SPT) phases. These are the phases classified by the effective theory of the $G$ gauge field. Recently, Anton Kapustin \cite{K} proposed that bosonic SPT phases whose symmetry group $G$ contains an orientation reversing symmetry are classified by a certain $G$-equivariant cobordism group.

In this note, we analyze various operators in these effective gauge theories. We find that it is natural to include in the definition of these operators a framing of their support. This framing can cause fermionic or other interesting braiding and fusion behavior for these quasiparticles and quasistrings.

Our results give evidence for the proposed description in \cite{K} for the SPT phases whose effective gauge theories are not Dikjgraaf-Witten theories (with local coefficients) \cite{CGLW}. Among these should be a 4d bosonic SPT with only fermionic quasiparticles on the boundary. We show that the boundary quasiparticles in the proposed effective gauge theory for this phase are indeed all fermions, but some for different reasons than others! We also give a description of ``fermionic" quasistrings on the 4d boundary of the novel 5d phase.

The 4d topological order considered in this paper demonstrates some of the loop-like braiding statistics postulated recently in \cite{wanglevin}. It has a $\ZZ/2$ charge and a $\ZZ/2$ flux-loop which have mutual semionic statistics and which are both fermions, in a sense defined below. This is a gapped system which realizes the gravitational anomaly $w_2w_3$. This action functional is a cobordism invariant and in fact generates $\Omega^5_{SO}$. In particular, if this system is realized on $\CC\mathbb{P}^2$, then observables change sign under complex conjugation (a large diffeomorphism). This anomaly can be cancelled by introducing neutral fermions, but remains if we introduce charged fermions.

Let us consider as our symmetry just time-reversal $G = \ZZ_2^T$. If the gauge field for this symmetry has nontrivial holonomy around a loop in space-time, that loop is necessarily orientation-reversing, since the nontrivial element of the symmetry group is an orientation-reversing spacetime symmetry. This demonstrates that the topology of space-time determines the configuration of the gauge field. Throughout, we will comment on how this topology can be considered like a dynamical gauge field.

The reason Dijkgraaf-Witten theory does not describe some of these phases is that the time reversal gauge field may also have holonomy around surfaces and higher dimensional submanifolds. The entire configuration of the field is specified by the unoriented bordism class of space-time. This is specified by the Stiefel-Whitney numbers of space-time, so we can also say that the $\ZZ_2^T$ gauge field is the collection of the Stiefel-Whitney classes.

The examples of effective actions for such a field that we will discuss are
$$
\frac{1}{2}\int w_1^4 \in \RR/\ZZ
$$
$$
\frac{1}{2}\int w_2^2 \in \RR/\ZZ
$$
$$
\frac{1}{2}\int w_2 w_3 \in \RR/\ZZ,
$$
where $w_j$ is the $j$th Stiefel-Whitney class. The first two describe 4d SPT phases, and the third describes a 5d SPT phase. The first is captured by the group cohomology classification (which only sees the 1-form part $w_1$ of the gauge field), while the second two are not. The second action is the one with all-fermion topological order, and the third has a ``fermionic" quasistring. 

Since these describe invertible field theories, we also think about them as describing anomalies in one less dimension. The first describes anomalous time-reversal symmetry, while the second two are more like gravitational anomalies: they cannot be canceled even if one breaks time reversal symmetry (which in the effective field theory corresponds to setting the 1-form part $w_1 = 0$). Put in the language of cobordisms, the second two represent non-trivial classes in $\Omega^*_{SO}$ as well as $\Omega^*_O$. Note that the second action becomes the same as a gravitational theta angle of $\pi$ in $\Omega^4_{SO}$, so it is continuously connected to a trivial action after breaking $T$-reversal. We will have more to say about this in future work.

Gauge transformations of the gauge field are space-time bordisms, which since the Stiefel-Whitney classes are bordism-invariants amounts to shifting the $w_j$ by exact $\ZZ_2$-cocycles. If we consider a space-time with boundary, the actions written above are no longer gauge invariant mod $\ZZ$.

Let us consider the second example. If we shift $w_2 \mapsto w_2 + \delta h$, then
$$
S \mapsto S + \frac{1}{2} \int_{\partial X} h \delta h.
$$
something also not gauge-invariant needs to live on the boundary to cancel this variation. The boundary theory we consider is
$$
S_{\rm all\ fermion} = \frac{1}{2}\int a \delta b + (a+b)w_2,
$$
where $a,b$ are integral 1-cochains representing $\ZZ_2$ gauge fields living on the 3d boundary. The action is invariant mod $\ZZ$ under the boundary gauge transformations
$$
a \mapsto a + \delta f + 2\alpha
$$
$$
b \mapsto b + \delta g + 2\beta.
$$
 Under the bulk gauge transformation parametrized by $h$, $a$ and $b$ transform as
$$
a \mapsto a+h
$$
$$
b \mapsto b+h.
$$
The action is not invariant under this transformation. It transforms as
$$
S_{\rm all \ fermion} \mapsto S_{\rm all\ fermion} + \frac{1}{2}\int h\delta h,
$$
cancelling the boundary variation of the bulk theory.

The equations of motion for $a$ and $b$ in this boundary theory are
$$
\delta a =\delta b =w_2.
$$
This implies that the ordinary Wilson loops
$$
\exp(i\pi\int_\gamma a)
$$
are not topological. The correct definition of the $a$ quasiparticle must be something else.

\section{The Stiefel-Whitney class $w_2$ and fermionic particles}

Let us consider the quasiparticle
$$
\exp(i\pi \int_\gamma a)
$$
where $\delta a =w_2$. Note that since $a$ is integer-valued, this is just $\pm1$. It is well known that $w_2$ obstructs the existence of a spin structure. A spin structure is precisely what we need to define a neutral spinor. Note that it is easier to define charged spinors since the gauge field may have some curvature cancelling the $w_2$ obstruction.

The yoga of obstruction theory is that a trivialization of the obstruction--eg. $\delta a = w_2$--is the same as the sort of structure that is obstructed, ie. we should think of $a$ as a spin structure. Then it is well-known (see eg. \cite{scorpan}) that a spin structure is the same as an assignment of $\pm 1$ to framed curves which flips signs when the framing is rotated by $2\pi$. Let us therefore frame $\gamma$, writing $\hat\gamma$, and define the framed Wilson line
$$
\exp(i\pi \int_{\hat \gamma} a)
$$
as this $\pm 1$. Concretely, we can use the framing to make a nearby curve $\gamma'$ and write
$$
\exp(i\pi \int_{\hat \gamma} a) = (-1)^{{\rm link}(\gamma,\gamma')} \exp(i\pi\int_\gamma a).
$$

Let us see how this framed Wilson line describes a fermionic quasiparticle. From the push-off formula we see that a $2\pi$ rotation of the quasiparticle gives a minus sign by increasing the linking number by one.
\begin{center}
\begin{tikzpicture}
\node (pic) {\includegraphics[width=0.4\textwidth]{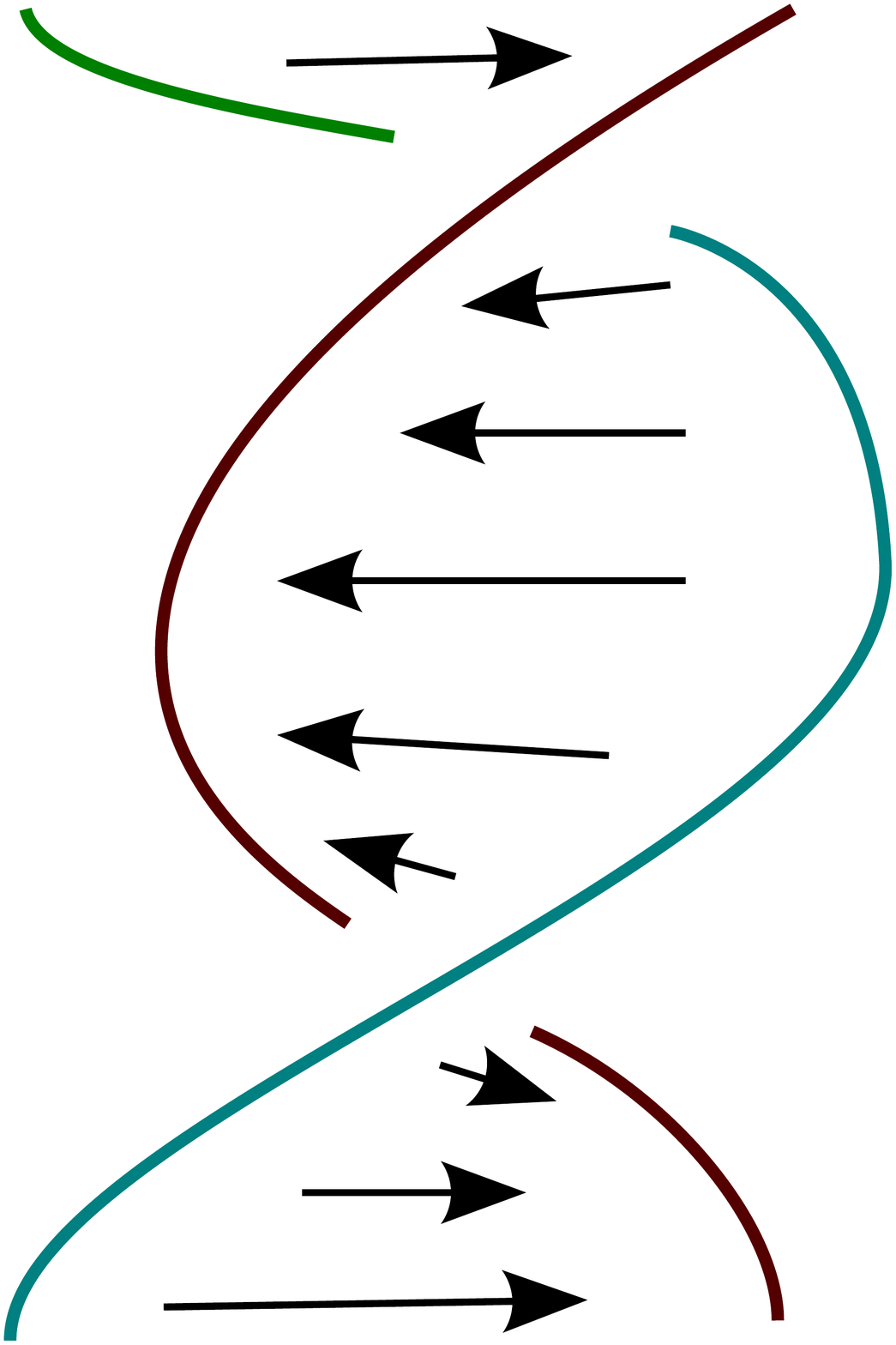}};
\node at (1.7,0) {$\gamma$};
\node at (-1.7,0.2) {$\gamma'$};
\end{tikzpicture}
\end{center}
The framing also causes fermionic braiding statistics, which we can see by creating a particle-antiparticle pair, braiding them, and then annihilating.
\begin{center}
\begin{tikzpicture}
\node (pic) {\includegraphics[width=0.25\textwidth]{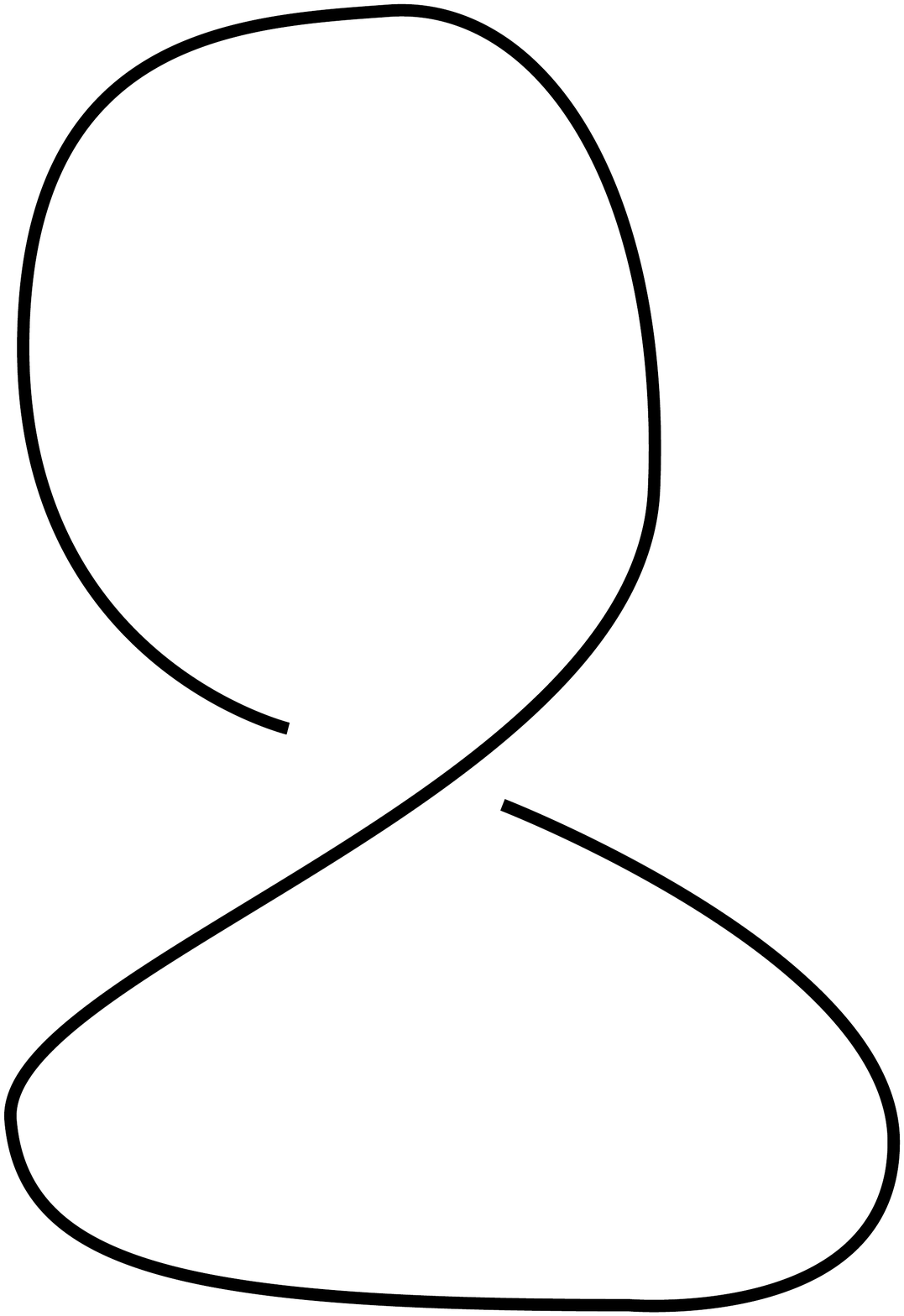}};
\node at (2,0) {{\rm framing\ }$\otimes$};
\end{tikzpicture}
\end{center}
As indicated, in this picture the framing always points into the page. This forces $link(\gamma,\gamma')=1$, so the braiding phase is $-1$.

\section{Stiefel-Whitney electric and magnetic operators}

Before we move on to general framed Wilson operators, let us discuss the electric and magnetic operators associated to the Stiefel-Whitney classes.

Again we begin with $w_2$. We can use Poincar\'e duality to represent $w_2$ by a (possibly unorientable) codimension 2 submanifold $X_{w_2}$. The homology class of this submanifold carries the same data as the cohomology class of $w_2$. The definition of $w_2$ implies that we can define neutral fermions in the complement of this submanifold. Then $X_{w_2}$ acts as a magnetic surface operator defined so that any wavefunction changes by $(-1)^F$ around a loop linking $X_{w_2}$.

This gives us a way of understanding of how $w_2$ acts as an obstruction. Indeed, if $X_{w_2}$ is non-trivial in $\ZZ/2$ homology, then there is no way to consistently define the linking number mod 2. In other words, the fermion parity cannot be consistently defined.

A simpler situation occurs with $w_1$. The Poincar\'e dual is a codimension 1 hypersurface in the complement of which we can define a consistent orientation on spacetime, but which flips orientation as we traverse $X_{w_1}$. Thus, it acts as a time-reversal or single-direction-inversion domain wall. Orientations cannot be consistently defined in the case when one cannot consistently decide which side of $X_{w_1}$ one is on. For example, $X_{w_1}$ for the M\"obius band cuts the band into a rectangle, but one is always on both sides of the cut.

It is more interesting to consider $w_1^2 = Sq^1 w_1$. This is the obstruction to lifting $w_1$ to a $\ZZ/4$ valued cocycle, or equivalently lifting $T$ to an order 4 symmetry. Indeed, we can consider $X_{w_2}$ as a codimension 2 magnetic operator such that fields transform by $T^2$ around a loop linking it, ie. Kramers degenerate particles have boundary conditions around this codimension 2 submanifold twisted by a minus sign. In this situation, the obstruction is interpreted as an inability to consistently define which particles have $T^2 = 1$ and which have $T^2 = -1$.

We can also have more complicated magnetic operators corresponding to things like $w_3$. In four dimensions this is a magnetic line around which a linking fermionic worldsheet picks up a minus sign. We will say more about these fermionic strings below.

We find it interesting to think about introducing such magnetic operators into the path integral. This effectively changes the topology of our spacetime, so we can think about evaluating observables on non-trivial topologies that we have obtained by insertion of magnetic operators into a contractible space.

There are also electric operators, such as the $w_1$ line
$$
\exp(i \pi \int_\gamma w_1).
$$
This has a natural interpretation as a $T$-odd particle traveling along $\gamma$. There are surface operators
$$
\exp(i\pi \int_\Sigma w_2).
$$
This does not have a simple analogous description, but it is interesting to consider the case where the surface has boundary $\partial \Sigma = \gamma$. In this case, the $w_2$ surface is not gauge invariant under the transformations
$$
w_2 \mapsto w_2 + \delta h.
$$
It transforms by
$$
\exp(i\pi \int_\gamma h).
$$
Thus, if we choose some $a$ with $\delta a = w_2$ that transforms as $a \mapsto a +h$, then the composite operator
$$
\exp(i\pi \int_\Sigma w_2 + \pi i \int_\gamma a)
$$
is gauge invariant. As we discussed above, this means that the $w_2$ surface must end on a fermion. Note that the vector normal to $\gamma$ in $\Sigma$ defines a framing of $\gamma$ and the above composite operator is the same as our framed Wilson line.

We can give an electric interpretation to the obstruction now. If we want to define a fermionic Wilson line as a composite $w_2$ surface, we must choose a surface bounding $\gamma$. Different choices of surfaces can make operators that differ by the integral of $w_2$ about a closed surface. This is trivial precisely in the case that $w_2$ is exact.

Similarly we find that a $w_1^2$ surface must end on a particle with $T^2 = -1$ and a $w_3$ volume must end on a fermionic string. These will all be defined also with respect to a framing.

How can a magnetic operator end? This is only possible on the boundary of spacetime. The operator will end in some sort of boundary excitation, and because the Stiefel-Whitney classes of the bulk restrict to the Stiefel-Whitney classes of the boundary, this excitation will behave like the corresponding magnetic operator on the boundary.

For example, in the case with symmetry $U(1) \rtimes \ZZ/2^T$, we have one non-trivial action
$$
\frac{1}{2} \int_X w_1 \frac{F}{2\pi}.
$$
We can use Poincar\'e duality to rewrite this as an integral on the magnetic $w_1$ surface
$$
\frac{1}{2} \int_{X_{w_1}} \frac{F}{2\pi}.
$$
Thus, all the interesting properties of this phase can be described by saying the magnetic $w_1$ surface is decorated with a 2d theta angle of $\pi$! As explained in \cite{K2}, such an action supports a charge $1/2$ zero mode on the boundary. The $w_1$ surfaces are time-reversal domain walls, and we can consider allowing them to proliferate. The half-charged ends will become a half-charged deconfined excitation, and because it is the end of a $T$-reversal domain wall, the dual vortex will be $T$-odd.

For a more sophisticated example, in \cite{KTTW} it was shown that an order 8 fermionic phase with $T^2 = (-1)^F$ in 4d can be characterized by decorating the magnetic $w_1^2$ surface with the Kitaev chain, so this surface ends on the boundary in the worldine of a Majorana zero mode.

We find it interesting to consider electric-magnetic duality in this context. The electric operators are defined using the Stiefel-Whitney cocycles and can end on objects with interesting time reversal properties or statistics, while the magnetic operators are defined using the Poincar\'e duals to these Stiefel-Whitney classes.

\section{Framed wilson operators}

This section is rather mathematical, but is necessary to describe the general construction. The next has pictures of the fermionic string.

The description of a spin structure we want to use to define the framed Wilson fermion is an assignment of $\pm 1$ to framed curves which flips sign when we twist the framing by $2\pi$. Mathematically, we can phrase this as $\ZZ/2$ 1-cocycle $\eta$ on the oriented frame bundle $P$ over spacetime $X$ which assigns $-1$ to the loop in the fiber, $SO(d)$, where $d$ is the spacetime dimension (and one can use Lorentzian signature if one chooses).

We want to produce such an object from a $\ZZ/2$ 1-cochain $a$ on spacetime satisfying $\delta a = w_2$, where we have made some universal choice of cocycle representative of $w_2$. Since the pullback $\pi^* P$ of the frame bundle to the frame bundle itself has a tautological section, there is a canonical trivialization $t$ of $\pi^*w_2$. This turns out to be the cochain that assigns $-1$ to the nontrivial loop in the fiber $SO(d)$ and $+1$ to all other loops. Then
$$
\delta t + \delta \pi^* a = \pi^* w_2 + \pi^* \delta a = \pi^* w_2 + \pi^* w_2 = 0,
$$
so $t + \pi^* a$ is a $\ZZ/2$ 1-cocycle on $P$. One checks that it assigns $-1$ to all the fiber loops, so defines a spin structure in the sense we want.

In order to interpret the twisting move for other Stiefel-Whitney classes, we need a description of the corresponding canonical trivialization $t$.

Let us describe where the Stiefel-Whitney classes come from. For references, see \cite{MS} We consider the ``Stiefel manifold", $V_k(\RR^d)$, whose points are $k$-tuples of orthonormal vectors in $\RR^d$. $SO(d)$ acts on this vector-wise, so to the frame bundle we can associate a bundle of these things where $d$ is the dimension of $X$. The class $w_{d-k+1}$ is the basic obstruction to finding a section of this bundle, ie. to finding $k$ everywhere orthonormal vector fields. Thus, eg. if $w_d \neq 0$ then $X$ does not have a non-vanishing vector field.

The Stiefel manifold can be understand as an iterated sphere fibration. The choice of the first unit vector is a point on $S^{n-1}$. Then we must choose a unit vector on the hyperplane normal to that vector, which can be thought of as the tangent space to $S^{n-1}$ at that point. Thus, $V_k(\RR^d)$ is the $k$-tuply iterated unit tangent bundle of $S^{d-1}$.

Once we've chosen all but the last vector, our remaining choice is a point on $S^{d-k}$. Since this is the smallest sphere in the fibration, we conclude $\pi_{d-k} V_k(\RR^d) = \pi_{d-k}S^{d-k} = \ZZ$ and all the lower ones are zero. By Hurewicz, $H_{d-k}(V_k(\RR^d))$ is generated by this element. If we send the generator to $-1$, we get an element of $H^{d-k}(V_k(\RR^d),\ZZ_2)$. This element is essentially our $t$. Its trangression from the associated Stiefel bundle down to $X$ is $w_{d-k+1}$.

Our problem is really just to understand this homotopy generator. Let's consider $w_2$. The relevant Stiefel manifold is $V_{d-1}(\RR^d)$ and we want to understand the fundamental group. A loop here can be thought of as a curve in $\RR^d$ with framed normal bundle (which is rank $d-1$). The generator, by the description above is given by fixing all but the last vector along the curve and letting the last vector make a rotation
\begin{center}
\begin{tikzpicture}
\node (pic) {\includegraphics[width=0.85\textwidth]{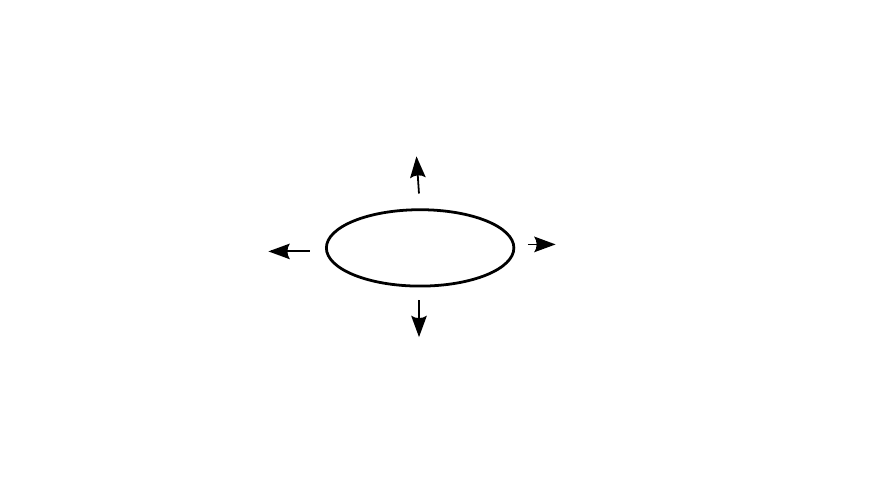}};
\node at (-4,-2.5) {Notice that we can pull this apart};
\end{tikzpicture}
\end{center}
\begin{center}
\begin{tikzpicture}
\node (pic) {\includegraphics[width=0.55\textwidth]{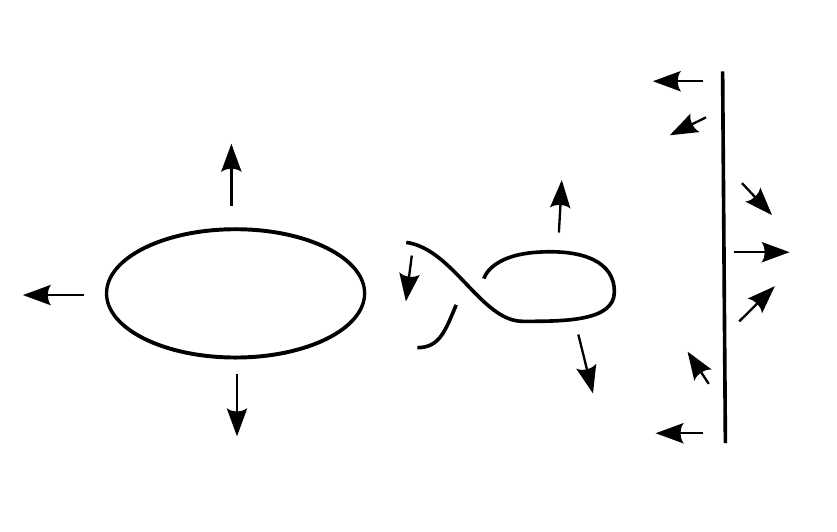}};
\end{tikzpicture}
\end{center}
Now we see the connection with the description earlier: this homotopy element is the twist!

Recall that a spin structure assigns $\pm 1$ to framed curves and the sign flips when the framing is twisted. Now we have the generalization.
\begin{thm}
A $w_{m+1}$-structure assigns $\pm 1$ to (normal) framed $m$-folds and the sign flips when the framing is twisted, where the twist is homologous in the frame bundle to summing with the generator of $\pi_m V_{d-m}(\RR^d)$.
\end{thm}

\section{$w_3$ and fermionic strings}

Finally we turn to the novel 5d SPT with effective Lagrangian $w_2 w_3$. The boundary 4d theory for this action is \cite{K}
$$
S_\partial = \frac{1}{2} \int a\delta b +a w_3 + bw_2,
$$
where $a$ is an integral 2-cochain and $b$ is an integral 1-cochain. The equations of motion are
$$
\delta a =w_3
$$
$$
\delta b =w_2.
$$
As we've seen above, this implies that the $b$ quasiparticle is a fermion. The electric operator for $a$ however is something new: a quasistring. Let us try to determine its braiding behavior.

For the $w_3$-structure $a$ (in four dimensions), we are talking about $V_2(\RR^4)$ and framed surfaces. The homotopy generator looks like the normal field of a 2-sphere in $\RR^3$ if we don't draw the fourth direction. Luckily the fourth direction is sort of boring. The 2-sphere is totally normal to it and the second vector of the normal frame is constant in that direction.

The interesting thing happens when we try to unfurl this picture as we did for spin structures. We need to make a tube do this. The twist rolls the framing around the string, like the natural rotation of a smoke ring. Here's what it looks like as a sphere:
\begin{center}
\begin{tikzpicture}
\node (pic) {\includegraphics[width=0.55\textwidth]{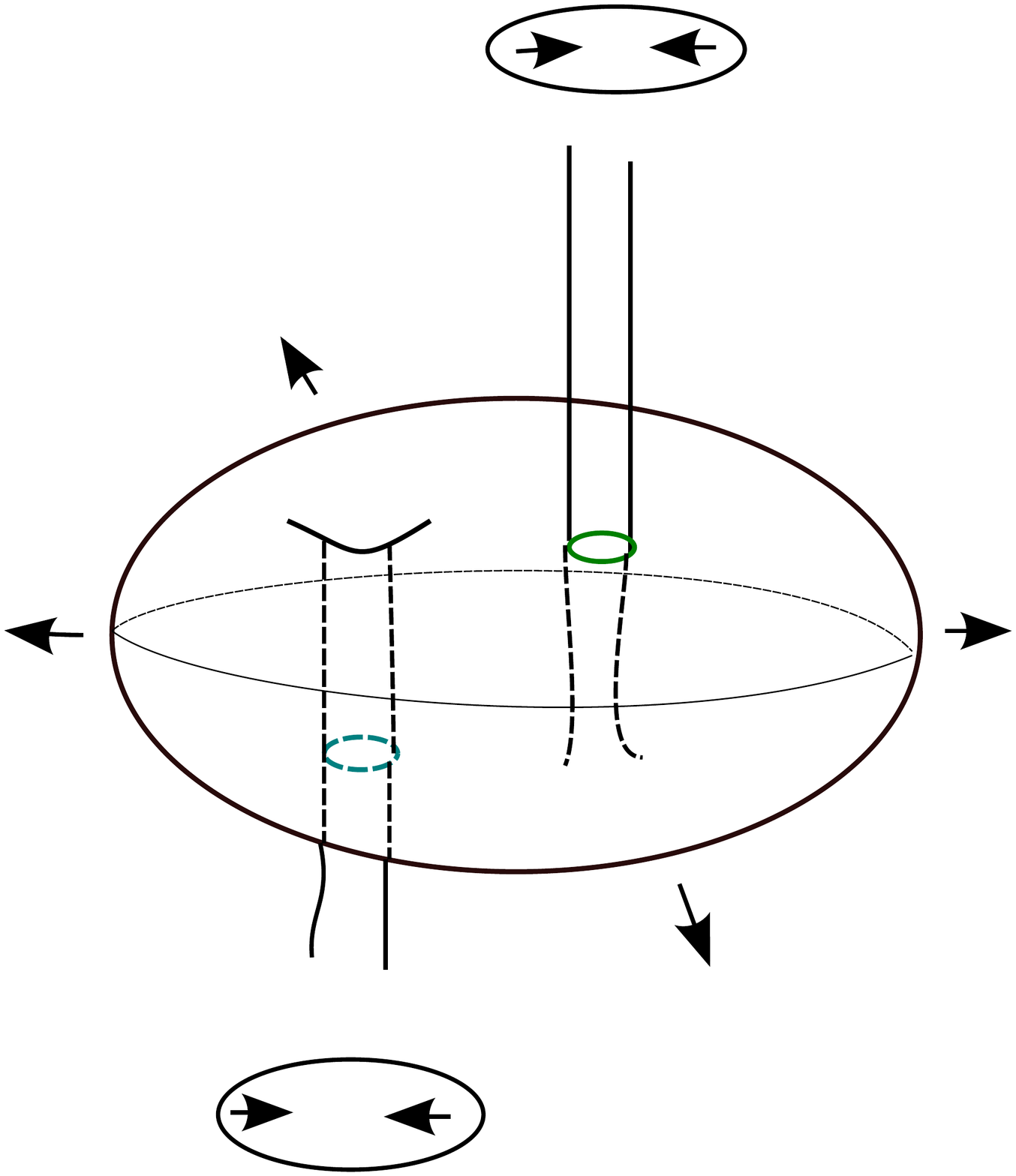}};
\end{tikzpicture}
\end{center}
(You can see in this picture the 2-sphere being thought of as a loop of loops.) There is some self-intersection in this picture, which must be resolved by some motion in the fourth dimension. We leave imagining this to the reader.

We can define $F$ now in a very similar way to the fermion case when the surface $\Sigma$ is homologically trivial. We consider the push-off $\Sigma'$ along one of the vectors in the frame (it doesn't matter which). There is a linking number $link(\Sigma,\Sigma')$ that changes by one under the twist. We can thus write the Wilson surface
$$
\exp(i\int_\Sigma a)(-1)^{link(\Sigma,\Sigma')}.
$$

\begin{center}
\begin{tikzpicture}
\node (pic) at (0,-3) {\includegraphics[width=0.35\textwidth]{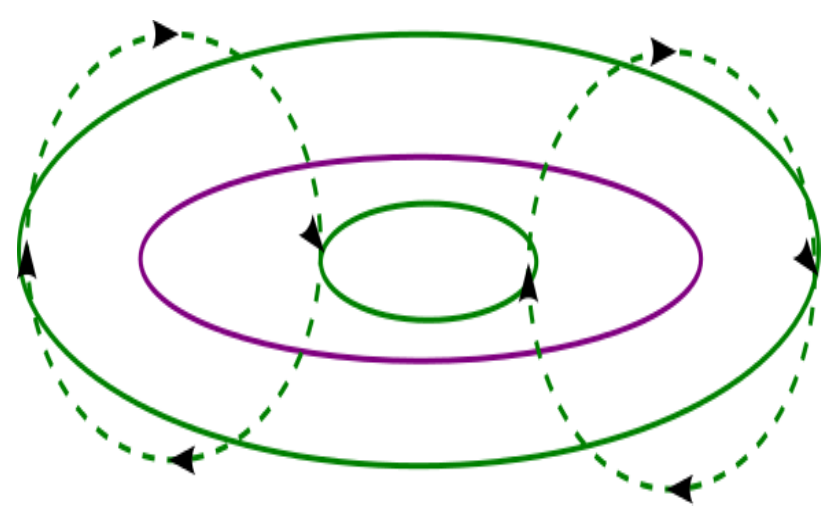}};
\end{tikzpicture}
\end{center}
(Color online) The nontrivial (full!) braiding for these strings looks like taking the first string (green) through the center of the second (purple) and around its outside, tracing a torus around the second string. This is the same two-loop braiding that was recently considered in \cite{wanglevin}.

Like the all-fermion topological order, this system has a gravitational anomaly (it is non-trivial even after breaking $T$ symmetry). Usually a gravitational anomaly is thought of as a non-trivial transformation rule under diffeomorphisms, so let us touch on this. Knowing the anomaly is $w_2w_3$, the path integral measure transforms by a phase
$$
\exp\big[i\pi\int_{X^S} w_2w_3\big],
$$
where $X^S$ is the mapping torus of a diffeomorphism $S$ of $X$. Note that if $S$ is isotopic to the identity, then $X^S$ is diffeomorphic to $X\times S^1$, which is the boundary of $X \times D$. Since $w_2w_3$ is a cobordism invariant,
$$
\int_{X\times S^1} w_2w_3 = 0.
$$
Thus, only *large* diffeomorphisms are involved in this anomaly. This makes sense, since in general we expect gapped systems have only global anomalies.

An important example that was also considered (for different but perhaps related reasons) in \cite{KGS} is $X = \CC\PP^2$. It can be shown that the mapping torus of complex conjugation generates $\Omega_5^O$, so in particular the path integral measure changes by $-1$ if the system has this anomaly.

Note that this can be cancelled by introducing neutral fermions, since these can form a bound-state with the $a$ quasiparticle. Then we're left with a system having a bosonic Z/2 charge, fermionic Z/2 flux, and neutral fermion. This system is not anomalous. However, if we introduce charged fermions, for example on $\CC\PP^2$, then there is still an anomaly since the gauge curvature must satisfy $\frac{F}{2\pi} = w_2 \mod 2$, so complex conjugation reverses its sign. There is no way for it to smoothly extend to the mapping torus. Note it is impossible to have neutral fermions on $\CC\PP^2$.

\section{QED with Fermionic Monopoles and 4d Gravitational Anomaly}

In this section we discuss an example of a system realizing the 4d global gravitational anomaly. This system is QED with a fermionic electron and a fermionic monopole. This system was also considered in \cite{WPS} who discussed what happens if you give this system a boundary. They found this is only possible with the introduction of neutral fermions. Here we explain this result by showing that this system has a gravitational anomaly (which we showed above can be cancelled by introducing neutral fermions).

Let $A$ denote the electromagnetic gauge field. We normalize the action so that it is an element of $\RR/\ZZ$. We will argue that to make the monopole fermionic, one must introduce a term
\begin{equation}\label{monopoleterm}
\frac{1}{2} \int_X w_2 F/2\pi.
\end{equation}
Here $w_2$ means an integer lift of the 2nd Stiefel-Whitney class. On an oriented, closed 4-manifold, this term actually equals
$$
\frac{1}{2} \int_X \frac{F}{2\pi} \wedge \frac{F}{2\pi},
$$
ie. we have turned on a theta angle of $\pi$ (but there is a subtle difference on unorientable manifolds).

Once this is proved, we condense Cooper pairs, producing a gapped phase with long range TQFT
$$
\frac{1}{2} \int_X a\delta b + \frac{1}{4} \int_X w_2 \delta a.
$$
Here $a$ is an integer valued 1-cochain which is closed mod 2. This represents the residual $\ZZ/2$ gauge field. It is related to $A$ after Higgsing by $a/2 = A \mod \ZZ$. Meanwhile, $b$ is an integer valued 2-cochain, closed mod 2, which is dual to the condensate (the term \eqref{monopoleterm} does not obstruct duality). In particular, the Wilson surface
$$
\exp(i\pi \int_\Sigma b)
$$
represents an insertion of a $\pi$-flux along the worldsheet $\Sigma$, which can be checked noting that it has semionic statistics with the $\ZZ/2$ charge (the electron).

Actually, to enforce fermionic statistics for the electron we must also add another term, so the final action is
$$
\frac{1}{2} \int_X a\delta b + \frac{1}{4} \int_X w_2 \delta a + \frac{1}{2} \int_X w_2 b.
$$
The magic is that \eqref{monopoleterm} can be integrated by parts in the Higgsed theory
$$
\frac{1}{4} \int_X w_2 \delta a = \frac{1}{2} \int_X \frac{\delta w_2}{2} a = \frac{1}{2} \int_X w_3 a,
$$
where we have used $Sq^1w_2 = w_3 + w_1 w_2$ and $w_1 = 0$ on an oriented manifold. This shows that the $\pi$-flux is a fermionic string, and we have shown that this all-fermion statistics has a gravitational anomaly with anomaly theory the cobordism TQFT with action $w_2w_3$. 

For now let us show that the term $\eqref{monopoleterm}$ produces a fermionic monopole. Let us consider the surface operator
$$
\exp\big[i\pi\int_\Sigma w_2\big].
$$
If this surface has boundary $\partial \Sigma = \gamma$, then this operator is not gauge invariant, but must be cancelled by an operator supported along $\gamma$. If we are to use an electric operator $\frac{1}{2}\int_\gamma c$, the gauge invariance condition is $dc = w_2$. In other words, the boundary particle must be a fermion. This argument shows any fermion will do, since all operators can be put in this electric form. Note that the surface $\Sigma$ can be thought of as defining a framing of $\gamma$.

Now consider an action containing the term \eqref{monopoleterm}. Let us hollow out a tube $T$ containing $\gamma$. Then under a gauge transformation $w_2 \mapsto w_2 + df$ the surface operator transforms as
$$
\exp\big[i\pi \int_{\gamma} f + i\pi \int_{\partial T} F f\big].
$$
We can write $\partial T = S^2 \times \gamma$ and if the sphere is small enough the second integral splits as a product
$$
i\pi\int_\gamma f \int_{S^2}F.
$$
From this we see that if $\int_{S^2} F = 1$, then this surface operator is gauge invariant. This is precisely the prescription for placing a magnetic monopole along $\gamma$. Thus, we've shown the $w_2$ surface can end on a magnetic monopole. Our argument above then implies the monopole is a fermion.

To finish the argument, we should show that the monopole was a boson to begin with. This amounts to the observation that without the extra term, there is no way the monopole could have been an end for the $w_2$ surface.

\section*{Acknowledgements}
I am grateful to Anton Kapustin for a collaboration on a related project and for sharing his many insights. I am also grateful to Xie Chen, Ashvin Vishwanath, and Andre Henriques for discussions. This paper also owes a lot of influence to John Baez and his wonderful blog.

\end{document}